\documentclass[a4paper,11pt]{article}
\usepackage{graphicx}
\usepackage{amsmath}

\input{tcilatex}

\begin{document}

\date{\today}

{\large \textbf{Experimental Evidence on Non-Applicability of the Standard
Retardation Condition to Bound Magnetic Fields and on New Generalized
Biot-Savart Law}}

\bigskip \bigskip

$\;\;\;\;\;\;\;\;\;\;$\textbf{A.L. Kholmetskii}

\bigskip

$\;\;\;\;\;\;\;\;\;\;$\textit{Department of Physics, Belorussian State
University}

\textit{$\;\;\;\;\;\;\;\;\;\;$F. Skorina Avenue 4, 220080, Minsk, Belorussia
}

\bigskip

$\;\;\;\;\;\;\;\;\;\;$\textbf{O.V. Missevitch}

\bigskip

$\;\;\;\;\;\;\;\;\;\;$\textit{Institute of Nuclear Problems, Belorussian
State University}

\textit{$\;\;\;\;\;\;\;\;\;\;$Bobruyskaya 11, 220050, Minsk, Belorussia}

\bigskip

$\;\;\;\;\;\;\;\;\;\;$\textbf{R. Smirnov-Rueda}

\bigskip

$\;\;\;\;\;\;\;\;\;\;$\textit{Applied Mathematics Department, Faculty of
Mathematics}

\textit{$\;\;\;\;\;\;\;\;\;\;$Complutense University, 28040 Madrid, Spain}

\bigskip

$\;\;\;\;\;\;\;\;\;\;$\textbf{R.I. Tzontchev, A.E. Chubykalo and I. Moreno}

\bigskip

$\;\;\;\;\;\;\;\;\;\;$\textit{Faculty of Physics, University of Zacatecas}

\textit{$\;\;\;\;\;\;\;\;\;\;$C-580, Zacatecas 98068, ZAC., Mexico}

\bigskip

\begin{abstract}
In this work we made an analysis of the current status of velocity
dependent (bound) field components in the framework of the standard
electromagnetic theory. Preliminary discussions of the structure of
the magnetic field due to an idealized oscillating magnetic dipole
provided us with the quantitative insights on the relative
contribution of velocity dependent (bound) and acceleration
dependent (radiation) terms into the resultant magnetic field.
According to this analysis we defined the methodological scheme
based on a generalized (time-dependent) Biot-Savart law capable to
test the applicability of the standard retardation condition on
bound field components. In the second part of this work we made the
theoretical analysis of the finite size multi-section antennas,
confirming the validity of the methodological scheme conceived for
an idealized magnetic dipole. The use of multi-section antennas is
fully justified by a substantial rise of the ratio of
bound-to-radiation field strength. Finally, we effected numerical
calculations taking into account particular experimental settings
and compared them with experimentally obtained data that
unambiguously indicate on the non-applicability of the standard
retardation condition to bound magnetic fields. In addition,
experimental observations show a striking coincidence with the
predictions of a new generalized Biot-Savart law which implies the
spreading velocity of bound fields highly exceeding the velocity of
light.
\end{abstract}


\section{Introduction}

Nowadays, the standard classical electrodynamics is perceived as the
paradigm of a causal and deterministic classical field theory. The most
general approach to the calculation of electric and magnetic fields is
explicable in terms of the so-called Lienard-Wiechert potentials. Their
conventional form in the case of an arbitrary moving classical point charge
of magnitude $q$ is given as$^{\cite{jackson}-\cite{landau}}$:

\begin{equation}
\varphi =\frac{q}{4\pi \varepsilon _{0}}\left[ \frac{1}{R-\frac{\mathbf{%
u\cdot R}}{c}}\right] _{ret};\qquad \mathbf{A}=\frac{q}{4\pi \varepsilon _{0}%
}\left[ \frac{\frac{\mathbf{u}}{c}}{R-\frac{\mathbf{u\cdot R}}{c}}\right]
_{ret}  \label{uno}
\end{equation}
where the corresponding quantity in brackets means that it is evaluated at
the retarded time $t^{\prime }=t-R/c$; $R$ is the distance from the retarded
position of the charge to the point of observation and $\mathbf{u}$ is a
velocity of a charge at the instant of time $t^{\prime }$.

As our intention to make clear the main proposals of this work, we first
discuss the fundamental underlying structure of existing solutions. Electric
and magnetic field values $\mathbf{E}$ and $\mathbf{B}$ are related to
potentials (\ref{uno}) by the equations:

\begin{equation}
\mathbf{E=}-\mathbf{\nabla }\varphi -\frac{\partial \mathbf{A}}{\partial t}%
;\qquad \mathbf{B}=\mathbf{\nabla }\times \mathbf{A}  \label{dos}
\end{equation}

Straightforward application of operations (\ref{dos}) performed on (\ref{uno}%
) provides the standard expression for field strengths of an arbitrary
moving point charge$^{\cite{jackson}-\cite{landau}}$:

\begin{equation}
\mathbf{E=}\frac{q}{(4\pi \varepsilon _{0})}\left\{ \left[ \frac{(\mathbf{n}-%
\frac{\mathbf{u}}{c}\mathbf{)(}1-(\frac{\mathbf{u}}{c})^{2})}{k^{3}R^{2}}%
\right] _{ret}+\left[ \frac{\mathbf{n}}{k^{3}R}\times \left\{ \left( \mathbf{%
n}-\frac{\mathbf{u}}{c}\right) \times \frac{\mathbf{\dot{u}}}{c}\right\} %
\right] _{ret}\right\}  \label{tri}
\end{equation}
\begin{equation}
\mathbf{B}=\left[ \mathbf{n}\times \mathbf{E}\right]  \label{cuatro}
\end{equation}
here $\mathbf{n}=\frac{\mathbf{R}}{R}$ and $k=1-\frac{\mathbf{n\cdot u}}{c}$%
;.

Both electric $\mathbf{E}$ and magnetic $\mathbf{B}$ fields are composed of
two \textit{essentially} different parts:

\begin{equation}
\mathbf{E}=\mathbf{E}_{\mathbf{u}}+\mathbf{E}_{\mathbf{a}};\qquad \mathbf{B}=%
\mathbf{B}_{\mathbf{u}}+\mathbf{B}_{\mathbf{a}}  \label{pyat}
\end{equation}
where $\mathbf{B}_{\mathbf{u}}=\mathbf{n}\times \mathbf{E}_{\mathbf{u}}$ and
$\mathbf{B}_{\mathbf{a}}=\mathbf{n}\times \mathbf{E}_{\mathbf{a}}$.

The first terms $\mathbf{E}_{\mathbf{u}}$ and $\mathbf{B}_{\mathbf{u}}$ are
usually regarded as \textit{velocity fields} because they are independent of
acceleration $\mathbf{\dot{u}}$. The electric velocity field $\mathbf{E}_{%
\mathbf{u}}$ falls off as $R^{-2}$ that is the substantial feature of the
radial (longitudinal) field components associated with the static Coulomb
law. By this reason the velocity fields are also referred as \textit{bound
or longitudinal fields}. On the contrary, the second parts of solutions (\ref%
{tri})-(\ref{cuatro}), denoted as $\mathbf{E}_{\mathbf{a}}$ and $\mathbf{B}_{%
\mathbf{a}}$ are linear functions of acceleration, falling off proportional
to $R^{-1}$. These terms are responsible for the radiation and take no place
in the case of uniformly moving charge (as well as at rest), giving priority
to bound components in the formation of electromagnetic (EM) field. The
latter fact poses a question of considerable fundamental interest concerning
the nature and physical characteristics on bound fields. It implies a
fundamental question of whether the applicability of the standard
retardation condition to all components of electromagnetic field has solid
theoretical, methodological and empirical grounds. Put in other terms,
should be there an explicit empirical test of the applicability of the
retardation condition to bound field components?

Doubts about the soundness of the conventional treatment of bound fields
appear already when one studies the transition between the initial state of
rest (or uniform motion) and the subsequent state of acceleration of the
charge. The origin of this important insight resides in the belief that a
fully consistent approach to classical electrodynamics demands the
continuity of electromagnetic phenomena in all situations. This allowable
inference was recently explored in$^{\cite{chub1}}$. The mathematical
analysis of boundary-value problems for D'Alembert and Poisson's equations
showed that the present set of solutions to Maxwell's equations does not
ensure the continuity in transition between, on one hand, the general case
of an arbitrary moving charge and, on the other hand, the static or
quasi-static (uniform motion) limits. Independent approach implemented in$^{%
\cite{vilecco}}$ also unambiguously indicates that the transition between
two different states of uniform velocity of a charge via an intermediate
state of acceleration results in a type of discontinuity in functional form.

The peculiar historical background of the classical EM theory gives an
additional motivation to focus our attention on bound fields. The problem of
propagation of EM interactions was the crucial point in choosing appropriate
theoretical foundations for classical electrodynamics in the 19th century.
By the time Hertz began his experiments in providing evidence in favor of EM
waves in air, the set of fundamental solutions to Maxwell's equations did
not imply any explicit separation into bound and radiation field components.
Only Lorentz's modification of Maxwell's theory (1892) provided Lienard
(1898) and Wiechert (1901) with inhomogeneous wave equations. The retarded
solutions were found under the retardation constraint which, as it is
generally accepted, was just experimentally verified by Hertz in 1888$^{\cite%
{hertz}}$. It gave a rise to the fact that bound and radiation field
components of Lienard-Wiechert solutions propagate at the same rate.

Thus, in light of our discussion we can see that Hertz's experiments were
not especially thought-out to bring solid empirical grounds on the physical
characteristics of bound fields as well as to test the applicability of the
retardation constraint on them. In modern retrospective, this uncertainty in
respect to bound components might imply some sort of methodological
incompleteness in Hertz's experimental approach as it has been recently
inferred in$^{\cite{smir}}$ and, therefore, any attempt to provide a fuller
experimental information on bound fields acquires a fundamental importance
in the framework of the classical EM theory. As response to these conceptual
questions, in this work we propose \textit{an experimental approach} to
velocity (bound) fields to test the applicability of the standard
retardation constraint to bound field components.

Recently, Hertz's-type experiment has already been attempted$^{\cite%
{tzontchev}}$ in order to remove above-mentioned uncertainties in respect to
electric bound field components. However, no explicit information was
obtained as regards to the nature of electric bound fields and the front of
the detected signal apparently propagated at the light velocity as it is
assumed in the standard EM theory. However, the subsequent analysis showed
that there was no preliminary theoretical estimation of the relative
contribution of bound and radiation components into the final signal. This
estimation is crucial if one takes into account that EM radiation terms fall
off as $R^{-1}$ and at large distances predominate, determining the time
shift of the signal front measured by oscilloscope. Additionally, the chosen
spheroid shape of the receiving antenna used in experimental measurements
made it especially unsensitive to the direction of the propagating electric
field.

There is also another experimental difficulty connected to the estimation of
the relative contribution of bound components when the radiation is very
intensive, i.e. for frequencies $\omega >10^{8}$ $s^{-1}$. In fact, the
intensity of EM radiation rapidly increases with $\omega $, hence hindering
the sensitivity of measuring devices in respect to bound fields that fall
off as $R^{-2}$. Contrarily, for the low frequency radiation the time
resolution of registration systems decreases considerably and results
insufficient to measure any EM signal with due precision. Without such
preliminary discussion and analysis, the experimental approach effected in$^{%
\cite{tzontchev}}$ looks rather unreliable. Put in other terms, one can not
avoid the preliminary calculation of an optimal range of variations of $R$
and $\omega $ in which the experimental treatment of bound fields might be
attended with expected precision.

In addition to the feebleness of the effects produces by bound fields that
fall off as $R^{-2}$, there are also several technical circumstances that
make direct observation of bound fields an extremely difficult task. All
known radiation devices (antennas etc) are designed to enhance the radiation
to the utmost, i.e. to reach the maximum value of the ratio of
radiation-to-bound field strength. In other words, the use of radiation
devices in their traditional form can be rather counterproductive and,
therefore, there should be a radically different experimental approach based
on a substantial rise of the ratio of bound-to-radiation field strength.

As follows, we shall present an experimental scheme that is methodologically
conceived to surmount all above-mentioned technical difficulties and to
provide unambiguous explicit information on bound magnetic components. The
main idea of our method is based on the analysis of EM field in the plane of
the loop antenna with oscillating current. Section $2$ is devoted to the
underlying idea of how the spreading velocity of bound magnetic fields can
be measured in an idealized case. In Section $3$ a methodological approach
will be described in more details as concerns a particular experimental
set-up (described in Section 4) which implements multi-section type of
antennas in order to increase the ratio of bound-to-radiation field
strenght. Finally, Section 5 presents results of numerical calculations and
their comparison with experimental data.


\section{The structure of EM field due to an idealized magnetic dipole:
introduction into methodology of experimental approach}

This Section will be devoted to the illustration of the background of the
consistent methodological approach to empirical identification of bound
magnetic fields. By analogy with the retarded vector potential $\mathbf{A}$,
the magnetic field $\mathbf{B}$ can be determined directly by the standard
expression$^{\cite{jefim}-\cite{rosser}}$:

\begin{equation}
B(R,t)=\frac{1}{4\pi \varepsilon _{0}c^{2}}\int \left[ \frac{\mathbf{\nabla }%
_{s}\times \mathbf{J}(r_{s})}{R}\right] dV_{s}  \label{seis}
\end{equation}
as the retarded solution of the corresponding D'Alembert equation:

\begin{equation}
\Delta \mathbf{B}-\varepsilon _{0}\mu _{0}\frac{\partial ^{2}\mathbf{B}}{%
\partial t^{2}}=-\mu _{0}\mathbf{\nabla }\times \mathbf{J}  \label{siete}
\end{equation}
where $\mathbf{J}$ is the conduction current density; $\mathbf{B}=\mathbf{%
\nabla }\times \mathbf{A}$ and the quantity placed inside the square bracket
in Eq. (\ref{seis}) is measured at the retarded time $t^{\prime }=t-R/c$.

In the low velocity relativistic limit (i.e. when the velocity of charges is
much lower than $c$ but the effect of retardation is still taken into
account)$^{\cite{jefim}-\cite{rosser}}$, the general solution (\ref{seis})
can be presented in an equivalent form of the line integral as the
generalized (time-dependent) Biot-Savart law$^{\cite{jefim}}$:

\begin{equation}
\mathbf{B}=\mathbf{B}_{\mathbf{u}}+\mathbf{B}_{\mathbf{a}}=\frac{1}{4\pi
\varepsilon _{0}c^{2}}\oint\limits_{\Gamma }\left\{ \frac{\left[ \NEG{I}%
\right] }{R^{2}}+\frac{\left[ \dot{I}\right] }{cR}\right\} \mathbf{k}\times
\mathbf{n}dl  \label{ocho}
\end{equation}%
where $\mathbf{n}=\mathbf{R}/R$; $\mathbf{I}$ is the conduction current; $%
\mathbf{k}$ is the unit vector in the direction of $\mathbf{I}$, i.e. $%
\mathbf{I=}I\mathbf{k}$ and $dl$ is an infinitesimal element of the loop
line $\Gamma $.

The generalized (time-dependent) Eq. (\ref{ocho}) is exact when the
centripetal acceleration of the moving charges is ignored (low velocity
approximation) and, as it is demonstrated in$^{\cite{rosser}}$, can be
derived directly, starting with (\ref{uno}) for retarded potentials $\varphi
$ and $\mathbf{A}$. As a consequence, the Eq. (\ref{ocho}) has a general
applicability to usual electric circuits since the velocity of conduction
electrons is always much lower than $c$. The first term on the right hand
side of expression (\ref{ocho}) is the classical integral form of the
Biot-Savart law when the current $I$ in the circuit $\Gamma $ is steady
(i.e. $\dot{I}=0$).

The velocity dependent (bound) component $\mathbf{B}_{\mathbf{u}}$ arises
from the term proportional to $\left[ I\right] $ whereas the acceleration
field $\mathbf{B}_{\mathbf{a}}$ is due to the $\left[ \dot{I}\right] $ term.
Traditionally, in the general (time-dependent) case solutions to
corresponding D'Alembert equations are found under the retardation
constraint applied to the whole electromagnetic field, hence providing the
same spreading velocity to bound and radiation fields as it takes place in
the standard Eq. (\ref{ocho}). In view of overwhelming empirical evidences
that radiation components propagate at the velocity of light $c$, we keep
the standard retardation constraint unmodified in respect to acceleration
fields $\mathbf{B}_{\mathbf{a}}$. Contrarily, we assume that there is no
explicit empirical evidence on the way how bound fields propagate in empty
space. Using the underlying structure of the Eq. (\ref{ocho}), we are now in
a position to deal with bound and acceleration fields separately and to
explore a wide range of hypothetic retardation constraints in respect to
bound components, i.e. the situation when the spreading velocity of bound
fields $\mathbf{B}_{\mathbf{u}}$, denoted as $v$, varies continuously in the
range $v\geq c$. In the context of this hypothesis, Eq. (\ref{ocho})
acquires an alternative form:

\begin{equation}
\mathbf{B}=\mathbf{B}_{\mathbf{u}}+\mathbf{B}_{\mathbf{a}}=\frac{1}{4\pi
\varepsilon _{0}c^{2}}\int \left\{ \frac{\left[ I\right] _{v}}{R^{2}}+\frac{%
\left[ \dot{I}\right] }{cR}\right\} \mathbf{k}\times \mathbf{n}dl
\label{ochoa}
\end{equation}%
where the corresponding quantity in square bracket $\left[ I\right] _{v}$
means that it is evaluated at the different retarded time $t_{v}^{\prime
}=t-R/v$.

The introduction of the dissimilar approaches to the treatment of bound and
radiation fields adds to the difficulty of its justification in the
framework of the conventional classical electrodynamics. The theoretical
exploration of this case will conceivably imply the extension of foundations
and perhaps the rise of new concepts. One possible justification might be
based on the use of a two-parameter Lorentz-like gauge (so-called $v$-gauge)$%
^{\cite{jackson1}}$ which leaves original Maxwell's equations unmodified and
introduces an arbitrary propagational velocity $v\geq c$ for scalar
components of electric field (i.e. they can be identified with bound
fields). Recently, $v$-gauge approach is widely discussed as a possible
explanation of apparently superluminal group velocity of evanescent waves$^{%
\cite{tittel}}$. Even though the foundations of the hypothetic distinction
of propagational velocities of bound and radiation components remain at
present stage unclear, the use of Eq. (\ref{ochoa}) allows us to conceive a
general methodological scheme capable to test the applicability of the
standard retardation constraint to bound fields, i.e. the case $v=c$.
Certainly, any discrepancy between experimental observations and theoretical
predictions based on the condition $v=c$ would not automatically imply the
correctness of our hypothesis in form of the Eq. (\ref{ochoa}) but rather no
applicability of the standard retardation constraint to bound fields. In the
latter case $v\neq c$, the Eq. (\ref{ochoa}) can be taken as a first
approximation to make theoretical predictions that can be verified by
experiment.

For further convenience, let us resort to a certain idealization in our
attempt to use the basic equation (\ref{ochoa}) in a less cumbersome
representation, keeping untouched all fundamental features related to the
bound and radiation components. By these reasons, we first consider an
example of idealized oscillating magnetic dipole given in form of a loop
antenna with the radius $r$. It will be assumed that the magnetic dipole
moment $\mathbf{m}$ varies harmonically with time at the angular frequency $%
\omega $. Then the idealization implies two requirements: $(a)$ the radius
of the loop $r$ is considerably smaller than the distance $R$ from the
center of the antenna to the point of observation; $(b)$ the wavelength $%
\lambda $ of emitted EM radiation greatly exceeds the perimeter of the loop,
i.e. $\frac{r}{c}\ll \frac{1}{\omega }$. The second requirement means that
at a given frequency $\omega $ the magnitude of conduction current has
nearly the same value $I$ in all parts of the loop circuit at a present
instant of time $t$. This condition, which is usually regarded as an
approximation of \textit{quasi-stationary current}, gives a simple
relationship $\mathbf{m}=\Delta SI(t)\mathbf{z}$ between the magnetic dipole
moment $\mathbf{m}$ and the conduction current value $I(t)$, where $\Delta S$
is the area bounded by the loop and $\mathbf{z}$ is the unit vector
perpendicular to the plane of the loop. Notice also that the \textit{%
quasi-stationary current} approximation remains valid for bound components
in the whole range $v\geq c,$ i.e. $\frac{r}{v}\ll \frac{1}{\omega }$.

Having assumed the above idealizations, the alternative expression (\ref%
{ochoa}) results considerably simplified. By analogy with the standard
procedure$^{\cite{rosser}}$ (a reader interested in a full derivation can
found it in the Appendix), all contributions into the magnetic field in the
plane of an oscillating magnetic dipole are:

\begin{equation}
\mathbf{B}(R,t)=-\frac{\Delta S}{4\pi \varepsilon _{0}c^{2}}\left\{ \frac{%
\left[ \NEG{I}\right] _{v}}{R^{3}}+\frac{c}{v}\frac{\left[ \dot{I}\right]
_{v}}{cR^{2}}+\frac{\left[ \ddot{I}\right] }{c^{2}R}\right\} \mathbf{z}
\label{nueve}
\end{equation}
where $\Delta S$ is the area bounded by the loop of emitting antenna (EA).

If the standard retardation constraint $v=c$ is applied to all components,
Eq. (\ref{nueve}) takes the form obtained in the low velocity relativistic
limit$^{\cite{rosser},\cite{stratton}}$:

\begin{equation}
\mathbf{B}(R,t)=-\frac{\Delta S}{4\pi \varepsilon _{0}c^{2}}\left\{ \frac{%
\left[ I\right] }{R^{3}}+\frac{\left[ \dot{I}\right] }{cR^{2}}+\frac{\left[
\ddot{I}\right] }{c^{2}R}\right\} \mathbf{z}=-\frac{\mu _{0}}{4\pi }\left\{
\frac{\left[ \mathbf{m}\right] }{R^{3}}+\frac{\left[ \mathbf{\dot{m}}\right]
}{cR^{2}}+\frac{\left[ \mathbf{\ddot{m}}\right] }{c^{2}R}\right\}
\label{diez}
\end{equation}

The first and the second terms in \textit{rhs} of Eq. (\ref{nueve}) have the
common origin and arise from the bound component $\mathbf{B}_{\mathbf{u}}$.
The contribution proportional to $R^{-3}$ is a dynamic counterpart of a
steady magnetic field produced by a static magnetic dipole of magnitude $%
\mathbf{m}$. The $R^{-2}$-term is due to a finite size of the antenna loop
which stipulates the difference\ in the value of retarded time $t^{\prime
}=t-R/v$ for EM signals out-coming from different segments of the perimeter
of EA. In other words, a bound field perturbation emitted from the nearest
half-part of the loop arrives at the point of observation before an
equivalent signal from the farthest half-part. This retarded time shift
leads to the contribution proportional to $\dot{I}$ or $\mathbf{\dot{m}}$
(for a more detailed explanation, see Appendix). The last $R^{-1}$-term is
submitted to the standard retardation constraint $v=c$ and corresponds to
the magnetic dipole radiation which falls off as $R^{-1}$.

Let us now take into consideration a receiving loop antenna (RA) under the
set of approximations assumed earlier for the EA. Let us place both EA and
RA loops in the same plane $xy$ as it is shown in Fig.1. The origin of the
coordinate frame coincides with the center of the EA loop and $\mathbf{R}$
denotes the position-vector of the center of the RA loop. \ The generation
of an electromotive force (e.m.f.) $\varepsilon (t)$ is proportional to the
time variation of the magnetic field flux (which is empirically established
integral form of the Faraday law):

\begin{equation}
\varepsilon (t)=-\frac{d}{dt}\iint\limits_{S}\mathbf{B\cdot z}dS
\label{once1}
\end{equation}

The time variation of the magnetic field $\mathbf{B}(t)$ in the area of the
RA is determined by the frequency of oscillations $\omega $ of the
conduction current $I(t)$ in the circuit of the EA. If the sizes of both EA
and RA are similar in magnitude, then in the approximation of the
quasi-stationary current for the EA the wavelength $\lambda $ of the
electromagnetic radiation also greatly exceeds the radius of the RA ($\frac{r%
}{c}\ll \frac{1}{\omega }$). In other words, the magnetic field $\mathbf{B}$
has nearly the same strength in all parts of the surface $S$ bounded by the
loop of the RA, so that:

\begin{equation}
\varepsilon (t)=-\frac{dB}{dt}\iint\limits_{S}dS=\frac{(\Delta S)^{2}}{4\pi
\varepsilon _{0}c^{2}}\left\{ \frac{\left[ \dot{\NEG{I}}\right] _{v}}{R^{3}}+%
\frac{c}{v}\frac{\left[ \ddot{I}\right] _{v}}{cR^{2}}+\frac{\left[ \dddot{I}%
\right] }{c^{2}R}\right\}  \label{once}
\end{equation}
where $\Delta S=\iint\limits_{S}dS$ is the area bounded by the loop of the
RA and $B=\mathbf{B\cdot z}$ is the component of the magnetic field in $%
\mathbf{z}$-direction.

The second $R^{-2}$-term disappears in the strong limit $v=\infty $. In this
hypothetic case, there is no retardation for bound fields and at the point
of observation every perturbation emitted from the nearest half-part of the
loop will be counteracted by an equivalent signal from the farthest
half-part. Therefore, the \textit{in-plane} geometry shown in Fig.1 should
be sensitive to any variation of the parameter $v$, since it directly
affects the value of the second $R^{-2}$-term.

This fundamental feature allows us to formulate a methodology of a
consistent experimental approach to bound magnetic components. Obviously, as
it has already been inferred in the Introduction, any practical realization
of the above-considered scheme may be accompanied by a number of rather
delicate technical difficulties. So, in our approach to idealization of
experimental conditions, let us first assume that the current $I(t)$ in the
EA oscillates harmonically at the angular frequency $\omega $ as $%
I(t)=-I(0)\cos (\omega t)$. Then, an observer can measure the phase of the
e.m.f. $\varepsilon _{v}(t)$ induced in the RA and compare it with the phase
of some reference harmonic signal $\varepsilon _{r}$. In the following, we
shall show how it can be implemented.

In the approximation of harmonic conduction current in the EA, Eq. (\ref%
{nueve}) yields:

\begin{equation}
\varepsilon _{v}(t)=\frac{(\Delta S)^{2}I(0)\omega }{4\pi \varepsilon
_{0}c^{2}}\left\{ \frac{\sin \omega (t-R/v)}{R^{3}}+\frac{\omega \cos \omega
(t-R/v)}{vR^{2}}-\frac{\omega ^{2}\sin \omega (t-R/c)}{c^{2}R}\right\}
\label{doce}
\end{equation}
or using notations $\eta =\omega R/c$ and $\eta _{v}=\omega R/v$ it can be
written as:

\begin{equation}
\varepsilon _{v}(t)=\frac{(\Delta S)^{2}I(0)\omega }{4\pi \varepsilon
_{0}c^{2}R^{3}}\left\{ \sin (\omega t-\eta _{v})+\eta _{v}\cos (\omega
t-\eta _{v})-\eta ^{2}\sin (\omega t-\eta )\right\}  \label{trece}
\end{equation}

Thus, all contributions into the value of the induced e.m.f. $\varepsilon
_{v}(t)$ are functions of $R$, $\eta $ and $\eta _{v}$ but at large
distances $R$, the radiation term proportional to $\eta ^{2}\sin (\omega
t-\eta )$ predominates. Therefore, the phase of the radiation term $(\omega
t-\eta )$ can be taken for the definition of the reference signal $%
\varepsilon _{r}(t)$:

\begin{equation}
\varepsilon _{r}(t)=-\varepsilon _{r}(0)\sin (\omega t-\eta )  \label{trecea}
\end{equation}

If the range of variations $v\geq c$ is assumed, there are two natural limit
cases: $(1)$ $v=c$ and $(2)$ $v=\infty $. Considering the first limit, which
corresponds to the standard assumption, Eq. (\ref{trece}) takes the form:

\begin{equation}
\varepsilon _{v=c}(t)=\frac{(\Delta S)^{2}I(0)\omega }{4\pi \varepsilon
_{0}c^{2}R^{3}}\left\{ \sin (\omega t-\eta )+\eta \cos (\omega t-\eta )-\eta
^{2}\sin (\omega t-\eta )\right\}  \label{cator}
\end{equation}

The first subplot in Fig.$2$ shows the comparison between the signal $%
\varepsilon _{v=c}(t)$ induced in the RA and the reference signal $%
\varepsilon _{r}(t)$, according to their evolution with $R$ and $t$. In the
second subplot of Fig.$2$ we specify relative contributions of $R^{-3}$, $%
R^{-2}$ and $R^{-1}$-terms and their mutual harmonic phase shifts. It
explains why $\varepsilon _{v=c}(t)$ and $\varepsilon _{r}(t)$ are not
synchronized, i.e. there is always a finite time shift $\Delta t$ between
instants when both signals $\varepsilon _{v=c}(t)$ and $\varepsilon _{r}(t)$
cross a zero line. However, at large distances the radiation $R^{-1}$-term
predominates considerably so that $R^{-3}$, $R^{-2}$-terms can be referred
only as weak perturbations of the dominant $R^{-1}$-component. In other
words, the signal $\varepsilon _{v=c}(t)$ tends to take the harmonic form
proportional to $\eta ^{2}\sin (\omega t-\eta )$, hence forcing the time
shift $\Delta t$ to vanish at large distances (far field zone).

The same analysis is also applicable in the general case for the
whole range of variations of the parameter $v$, i.e. $v\geq c$. In
Fig.$3$ we show diagrams that correspond to the second hypothetic
limit $(v=\infty )$. At short distances, there is also clearly
visible finite time shift $\Delta t$ when both signals
$\varepsilon _{v=\infty }(t)$ and $\varepsilon _{r}(t)$ cross a
zero line. We remind here that in this limit case the contribution
proportional to $R^{-2}$ disappears so that only $R^{-3}$ and
$R^{-1}$-terms can be seen in the second sub-diagram of Fig.$3$.
Moreover, if at distances $R>c/\omega $ the first $R^{-3}$-term
results already negligible in comparison with $R^{-1}$-term (it
will be fulfilled in our experimental conditions) the variation of
the time shift $\Delta t(R)$ does not exhibit any oscillations as
it is shown in Fig.$4$. We have thus reached the conclusion that
the type of variation of $\Delta t(R)$ is very sensitive to the
particular value of the parameter $v$ and hence can be subjected
to the empirical test. Further on we shall refer this approach as
a \textit{zero crossing method}.

Summarizing the discussion of idealized antennas, we would like to emphasize
that the requirements imposed by this approximation are hardly available in
real experimental practice. On the one hand, the value of the e.m.f. $%
\varepsilon $ depends essentially on the radius $r$ of the loop antenna
because it is proportional to $(\Delta S)^{2}\sim r^{4}$. On the other hand,
in order to reach the criteria of the quasi-stationary current
approximation, one cannot decrease the radius $r$ without moderation, since
the precision of measurements critically depends on the intensity of EA and
RA signals. Therefore, there should be a balance between mutually excluding
requirements. In our attempt to reduce dependence on the radius of the loop $%
r$, we resorted to the help of so-called multi-section antennas. As we will
show in the next Section, the approximation of \textit{quasi-stationary
current} and the maximum ratio of bound-to-radiation field strength can be
easily implemented for finite size multi-section antennas (even in the range
of high frequencies $\omega $), keeping valid the approach based on the zero
crossing method.

\section{Methodology of experiment in the case of multi-section antennas}

In this Section we shall discuss the methodology of experimental
approach to bound fields in the case of multi-section emitting and
receiving antennas which constitutes a practical realization of
particular experimental scheme considered in Section 4. The EA and
RA are composed from four and two sections, respectively, as it is
shown in Fig.$5$. The arrows mean directions of conduction
currents at some given instant of time. Therefore, the symmetric
form of multi-section emitting and receiving antennas implies no
net current in radial directions. This property is important to
reduce considerably undesirable electric dipole radiation in the
case of EA and, in addition, to suppress any disturbing effect
produced by electric dipole radiation in RA. Put in other terms,
it provides us with a higher ratio of bound-to-radiation field
strength (note that standard antennas are designed to reach the
highest ratio of radiation-to-bound field strength in order to
enhance the radiation). This circumstance is of crucial importance
as concerns empirical observations of bound magnetic fields as
well as the precision of measurements.

In respect to the approximation of quasi-stationary current, the use of
multi-section antennas improves considerably the basic criteria $\frac{r}{c}%
\ll \frac{1}{\omega }$. In fact, all radial parts of\ each section make no
contribution in generation of magnetic fields (bound or radiation) due to
the cancelation of current flows in adjacent parts of neighbor sections. It
stands out the role played by the arc fragment of every section, being thus
the only part of the multi-section antenna capable to produce magnetic
fields. Since the resultant magnetic field is determined entirely as a
superposition of magnetic fields created by each section, then the
applicability of the approximation of quasi-stationary current resides on
its validity for each section. As a consequence, we can consider the
criteria for the quasi-stationary current approximation of a multi-section
antenna as:

\begin{equation}
\omega \ll \omega _{c}=\frac{nc}{r}  \label{quinc}
\end{equation}
where $\omega _{c}$ denotes the critical value of the frequency of
oscillations and $n$ means the number of sections.

Thus, the use of multi-section type of antennas was fully
justified in our approach by the above-exposed practical reasons.
Let us now describe a technical realization of the experimental
set-up as regards the emitting multi-section antenna (see
Fig.$6$). A fast high-voltage spark gap (SG) constitutes the
driving circuit of the EA and it is connected with the
multi-section emitting circuit via the blocking capacitor $C$
which also plays the role of the energy storage element. The whole
circuit, presented in Fig.$6$, can be viewed as $LC$-contour with
the proper frequency $\omega _{0}=1/\sqrt{LC}$, where $L$ is the
inductance of the EA. We also assume that the values of $L$, $C$
and the radius $r$ are given in the range of applicability of the
approximation of quasi-stationary current, i.e. $\omega _{0}\ll
\omega _{c}$. The high-voltage supply (HV) is connected to the EA
via the resistor $R_{f}$. The latter determines the time necessary
to charge the capacitor $C$ as well as the duration of the period
in a series involving successive charges and discharges.

The magnitude of the initial current flow in the EA depends on the
value of high-voltage $U$ and the discharge time $\tau $. The
technical realization described above and presented in Fig.$6$,
generally speaking, does not imply harmonic oscillations in the
circuit of the EA. As a consequence, further on
we shall assume the exponential character of the discharge process, i.e. $%
U(t)=U(0)\exp (-t/\tau )$ and the shape of the emitted signal will be
conditioned by the Kirchhoff equation for the emitting circuit:

\begin{equation}
U_{L}+U_{R_{e}}+U_{C}=U(0)e^{-t/\tau }  \label{shestn}
\end{equation}
where $R_{e}$ is the characteristic resistance of the EA.

Taking into account standard relationships such as $U_{L}=L\frac{dI}{dt}=L%
\frac{d^{2}Q}{dt^{2}}$; $U_{C}=\frac{Q}{C}$ (where $Q$ denotes the charge)
and neglecting the resistance $R_{e}$ (the approximation valid when forced
oscillations take place), the Kirchhoff equation (\ref{shestn}) takes the
form:

\begin{equation}
\frac{d^{2}Q}{dt^{2}}+\omega _{0}^{2}Q=Q_{0}\omega _{0}^{2}e^{-t/\tau }
\label{semn}
\end{equation}
where $Q_{0}=U(0)/C$.

If one chooses the initial conditions $Q=Q_{0}$; $\frac{dQ}{dt}=0$, the
solution to Eq. (\ref{semn}) can be written as:

\begin{equation}
Q(t)=\frac{Q_{0}}{1+\omega _{0}^{2}\tau ^{2}}(\cos \omega _{0}t+\omega
_{0}\tau \sin \omega _{0}t+\omega _{0}^{2}\tau ^{2}e^{-t/\tau })
\label{vosemn}
\end{equation}
providing also the time dependent magnitude of the conduction current $I(t)$:

\begin{equation}
I(t)=\frac{dQ}{dt}=\frac{Q_{0}\omega _{0}}{1+\omega _{0}^{2}\tau ^{2}}(-\sin
\omega _{0}t+\omega _{0}\tau \cos \omega _{0}t-\omega _{0}\tau e^{-t/\tau })
\label{devyatn}
\end{equation}
and its respective time derivatives:

\begin{equation}
\frac{dI}{dt}=\frac{Q_{0}\omega _{0}^{2}}{1+\omega _{0}^{2}\tau ^{2}}(-\cos
\omega _{0}t-\omega _{0}\tau \sin \omega _{0}t+e^{-t/\tau })  \label{veinte}
\end{equation}

\begin{equation}
\frac{d^{2}I}{dt^{2}}=\frac{Q_{0}\omega _{0}^{3}}{1+\omega _{0}^{2}\tau ^{2}}%
(\sin \omega _{0}t-\omega _{0}\tau \cos \omega _{0}t-\frac{1}{\omega
_{0}\tau }e^{-t/\tau })  \label{veinte1}
\end{equation}

We are now in a position to approach a numerical evaluation of the
output signal in the RA. Let the loops of the emitting and
receiving antennas belong to the same plane $xy$ as it is shown in
Fig.$7$. The general expression (\ref{ochoa}) gives the resultant
magnetic field $\mathbf{B}$ produced by the EA:

\begin{equation}
\mathbf{B}=\frac{1}{4\pi \varepsilon _{0}c^{2}}\oint\limits_{\Gamma }\left\{
\frac{\left[ I\right] _{v}}{R^{\prime 2}}+\frac{\left[ \dot{I}\right] }{%
cR^{\prime }}\right\} d\mathbf{l}\times \mathbf{n}_{R^{\prime }}
\label{veinte1a}
\end{equation}
where $R_{x}^{\prime }=R+r\cos \theta -r_{EA}\cos \varphi ;$ $R_{y}^{\prime
}=r\sin \theta -r_{EA}\sin \varphi ;$ $R_{z}^{\prime }=0$; $d\mathbf{l}$ is
an infinitesimal element of the emitting loop $\Gamma $ and $\mathbf{n}%
_{R^{\prime }}=\mathbf{R}^{\prime }/R^{\prime }$ is the unit vector.

The e.m.f. $\varepsilon (t)$ induced in the RA is due to the magnetic field $%
\mathbf{B}$ produced by the EA and can be calculated according to the
integral form of the Faraday induction law:

\begin{equation}
\varepsilon =-\frac{d}{dt}\int\limits_{S_{RA}}\mathbf{B}(R,t)\mathbf{\cdot }d%
\mathbf{S}=-\int\limits_{0}^{2\pi }\int\limits_{0}^{r_{RA}}\frac{dB_{z}(R,t)%
}{dt}rdrd\theta  \label{veinte3}
\end{equation}
where $S_{RA}$ and $r_{RA}$ are the area and the radius of the RA,
respectively.

Since we assume valid the approximation of quasi-stationary current, i.e. $%
\omega \ll \omega _{c}$, then at a given instant\ of time $t$ the conduction
current has nearly the same magnitude $I(t)$ in all parts of the loop of the
EA. Having this approximation in mind, we substitute (\ref{veinte1a}) in (%
\ref{veinte3}) and find that:

\begin{equation*}
\varepsilon =\frac{1}{4\pi \varepsilon _{0}c^{2}}\int\limits_{0}^{r_{RA}}%
\int\limits_{0}^{2\pi }\int\limits_{0}^{2\pi }\frac{\left[ \dot{I}\right]
_{v}}{R^{\prime 3}}\left( 1-\frac{r}{r_{EA}}\cos \left( \theta -\varphi
\right) -\frac{R}{r_{EA}}\cos \varphi \right) r_{EA}^{2}rdrd\theta d\varphi +
\end{equation*}

\begin{equation}
+\frac{1}{4\pi \varepsilon _{0}c^{2}}\int\limits_{0}^{r_{RA}}\int%
\limits_{0}^{2\pi }\int\limits_{0}^{2\pi }\frac{\left[ \ddot{I}\right] }{%
R^{\prime 2}}\left( 1-\frac{r}{r_{EA}}\cos \left( \theta -\varphi \right) -%
\frac{R}{r_{EA}}\cos \varphi \right) r_{EA}^{2}rdrd\theta d\varphi
\label{veinte4}
\end{equation}
where the square bracket $\left[ \dot{I}\right] _{v}$ means that the value
of $\dot{I}$ in (\ref{veinte}) is taken at the retarded time $t\rightarrow
t-R^{\prime }/v$\ whereas the term $\left[ \ddot{I}\right] $ implies the
substitution $t\rightarrow t-R^{\prime }/c$ in (\ref{veinte1}).

In order to bring our numerical evaluations as close as possible to the
actual realization of the experiment, we also have taken into consideration
the widths of the emitting and receiving antennas, denoted as $h_{EA}$ and $%
h_{RA}$ respectively. Thus, defining the current density per a unit length
in the EA as $I/h_{EA}$ we can rewrite (\ref{veinte4}):

\begin{equation*}
\varepsilon =\frac{1}{4\pi \varepsilon _{0}c^{2}h_{EA}}\int%
\limits_{0}^{r_{RA}}\int\limits_{0}^{2\pi }\int\limits_{0}^{2\pi
}\int\limits_{0}^{h_{RA}}\frac{\left[ \dot{I}\right] _{v}}{R^{\prime 3}}%
\left( 1-\frac{r}{r_{EA}}\cos \left( \theta -\varphi \right) -\frac{R}{r_{EA}%
}\cos \varphi \right) r_{EA}^{2}rdrd\theta d\varphi dh+
\end{equation*}

\begin{equation}
+\frac{1}{4\pi \varepsilon _{0}c^{2}h_{EA}}\int\limits_{0}^{r_{RA}}\int%
\limits_{0}^{2\pi }\int\limits_{0}^{2\pi }\int\limits_{0}^{h_{RA}}\frac{%
\left[ \ddot{I}\right] }{R^{\prime 2}}\left( 1-\frac{r}{r_{EA}}\cos \left(
\theta -\varphi \right) -\frac{R}{r_{EA}}\cos \varphi \right)
r_{EA}^{2}rdrd\theta d\varphi dh  \label{veinte5}
\end{equation}
where $R^{\prime }=\sqrt{R_{x}^{\prime 2}+R_{y}^{\prime 2}+h^{2}\text{.}}$

The form of the calculated e.m.f. $\varepsilon (t)$ depends on the parameter
$\tau $ which was introduced above as a parameter responsible for the
duration of discharge processes in the circuit of the EA. In addition, we
have taken into consideration the proper rise time of the RA ($\tau _{RA}$)
and the proper rise time of the oscilloscope ($\tau _{o}$) in the circuit of
the RA. It allowed us to use an effective rise time $\sqrt{\tau ^{2}+\tau
_{RA}^{2}+\tau _{o}^{2}}$ in order to make more realistic calculations
concerning the front of the e.m.f. $\varepsilon (t)$ signal. For numerical
calculations on base of the Eqs. (\ref{veinte})-(\ref{veinte5}) we used
\textit{Mathcad Professional 2000 software}. All necessary parameters as
regards to the particular experimental settings, will be described in the
next Section.

\section{Experimental settings}

The emitting multi-section antenna (see Fig.$5a$) was assembled
according to the following geometric parameters: \textit{(1)} the
radius $r=50mm$; \textit{(2)} the width $h_{EA}=50mm$ and
\textit{(3)} the gap between adjacent sections - $3mm$. The frame
of the EA was made of copper sheet with $1mm$ thickness. To
deliver a current to every section simultaneously, we
used four parallel cables, each one having the characteristic resistance of $%
50Ohms$ and the length of $20cm$. The inductance of the EA (all four
sections are connected in parallel) is $L=46\pm 1$ \textit{nH} . It was
measured by the impedance meter \textit{E7-14} (Russia) at frequency $10kHz$.

To charge the \textit{RC} circuit, we used a high-voltage supply of $U\leq
2kV$. The spark gap \textit{EPCOS B88069-X3820-S102, A71-H10X 1400V/10A} has
a discharge voltage of $1.4kV$. Estimated discharge time of the \textit{RC}
circuit is $3ns$. The capacitor \textit{C} has $22pF$, thus determining the
proper frequency $\omega _{0}=1/\sqrt{LC}=9.9\times 10^{8}$ $s^{-1}$.
However, the real measured frequency of current damped oscillations ($t\gg
\tau $) is $8.8\times 10^{8}$ $s^{-1}$. It turned out to be a bit lower than
the expected value $\omega _{0}$ due to the presence of small proper circuit
inductance and capacitance which are usually unknown. We found values $%
r=50cm $ and $h_{EA}=50$\ to be optimal for the size of the EA. In this
case, according to the definition (\ref{quinc}), the critical frequency $%
\omega _{c}$ is $2.4\times 10^{10}$ $s^{-1}$, hence assuring the validity of
the approximation of quasi-stationary current.

The capacitor $C$ and the resistor $R_{f}=5MOhms$ determine the duration of
the period between discharges $T_{d}\approx 0.1ms$ which is important to
exhibit the signal on the oscilloscope screen with due brightness. A
synchronizing time signal was generated by the Rogovski belt$^{^{\cite{rogov}%
}}$, placed between the output of the discharge circuit and the input of the
EA. The spark gap and the capacitor were kept inside the magnetic and
electric shielding. It is also worth noting that a relative variation in
amplitude of generated signals (caused by the instability of a threshold
voltage in SG discharges) did not exceed $5\%$ of the average value.

The geometrical parameters of the RA (see Fig.$5b$) are:
\textit{(1)} the radius $r=50mm$; \textit{(2)} the width
$h_{RA}=100mm$ and \textit{(3)} the gap between adjacent sections
- $3mm$. Both sections of this antenna are connected in parallel.
Their frames are also made of copper sheet with $1mm$
thickness. The inductance of the RA, measured by the impedance meter \textit{%
E7-14} (Russia) at frequency $10kHz$, is\textit{\ }$L=49\pm 1nH$. The proper
rise time of the RA (coupled to the cable of $50Ohms$) can be estimated as $%
\tau _{RA}=\frac{49nH}{50Ohm}\approx 1ns$. We found values $r=50cm$ and $%
h_{RA}=100mm$ to be optimal for the size of the RA. The corresponding
critical frequency $\omega _{c}=1.2\times 10^{10}$ $s^{-1}$ is still one
order of magnitude higher than the proper frequency of current oscillations $%
\omega _{0}$, keeping the validity of the approximation of the
quasi-stationary current for the RA.

To visualize signals generated in the RA, we used the oscilloscope \textit{%
C1-108} (Russia) with the time scale $1ns$ per division. The proper rise
time of the oscilloscope is $\tau _{o}\approx 1ns$ and the expected time
resolution is about $0.1ns$. Additionally, a maximal voltage sensitivity
available by the oscilloscope is $0.01V$ per division. Both antennas were
mounted on a wooden table and all metallic objects (with the capacity to
reflect EM radiation) were removed from the experimental installations at
distances exceeding $1.5m$. It assured to make observations in such a manner
that during the period of the first $10ns$, all EM perturbations received by
the RA could in no way be attributed to reflected fields.

\section{Experimental measurements}

We remind the reader that the \textit{zero crossing method} (see Section $2$%
) provides the type of variations $\Delta t(R)$ which are very
sensitive to the particular value of the parameter $v$ from the
range $v\geq c$ and, therefore, it should be considered as a
reliable empirical test of the applicability of the standard
retardation condition ($v=c$) to bound fields. In Fig.$8$ we show
the results of numerical calculations (for multi-section antennas)
on the base of the Eqs. (\ref{veinte})-(\ref{veinte5}) as regards
the assumption on the undefined parameter $v\geq c$ which was
introduced to represent the spreading velocity of bound field
components. Dot lines describe the time shift $\Delta t$
dependence on the distance $R$ between emitting and receiving
antennas at the retardation conditions $v=c$; $v=2c$ and $v=10c$.
The limit case $v=\infty $ we plotted in a continuous line. It
is worth emphasizing that the types of numerical predictions for $v=c$ and $%
v=\infty $ are qualitatively in agreement with the predictions
obtained in the case of an idealized magnetic dipole (see
Fig.$4$).

Now we are in a position to apply the \textit{zero crossing
method} and to subject theoretical predictions $\Delta t(R)$ to
empirical test. For this purpose, we placed the emitting and the
receiving antennas in parallel positions as it is shown in
Fig.$7$. Keeping the orientation of antennas unchanged, we varied
the distance between them in the range of $R=40\div 200cm$,
applying the step of $\Delta R=20cm$. At each position we
succeeded in producing detectable signals on the oscilloscope
screen. In order to implement the \textit{zero crossing method}
(discussed in Section 3), we measured the instant $t_{cross}$ when
the oscillating disturbance crosses zero line for the first time.
At short distances, the amplitude of the whole signal is large,
hence providing the maximum steepness to the front of the signal.
It allowed us to measure the value of $t_{cross}$ with due
precision (the estimated error was less than $0.1ns$).

To complete our measurements we had to obtain the information on the instant
$t_{cross}^{\prime }$ when the reference signal (i.e. the harmonics of the
type $\varepsilon _{r}=-\sin (\omega t-\omega R/c)$) crosses zero line for
the first time. Since it cannot be done in direct measurements (the EA does
not generate the reference signal alone), we proposed indirect but
nevertheless unambiguous determination of the value of $t_{cross}^{\prime }$%
. As it already has been stated, at large distances $R^{-3}$,
$R^{-2}$-terms result negligible in comparison with the radiation
$R^{-1}$-term. In fact, the numerical calculations based on Eq.
(\ref{veinte5}) indicated that at the distance of $R=200cm$
between the EA and RA, a relative contribution of both $R^{-3}$
and $R^{-2}$-terms into the final signal is less than $0.1\%$. It
means that at $R=200cm$ there is practically no difference between
the values of $t_{cross}$ and $t_{cross}^{\prime }$ (here we
remind a reader to see diagrams in Figs.$2$ and $3$). Thus, the
value of $t_{cross}^{\prime }$ once established at $R=200cm$, can
be easily recalculated for any $R<200cm$.
It provides us the reliable experimental evaluation of the difference $%
\Delta t=t_{cross}-t_{cross}^{\prime }$ at each $R$ from the range
$R=40\div 200cm$. A graphic visualization of the empirically found
dependence $\Delta t(R)$ we present in Fig.$8$ (black circles).
The unexpected behavior of the experimentally found dependence
$\Delta t(R)$ is strikingly evident and it is apparently at odd
with the predictions of the standard EM theory (the dot line
$v=c$).

\section{Conclusions}

Based on these results, one can conclude that empirical data unambiguously
indicate on the non-applicability of the standard retardation constraint ($%
v=c$) to bound fields. Since the standard classical electrodynamics is
generally acknowledged nowadays as the most orthodox example of a causal and
deterministic classical field theory, one can suppose that the standard
theory may contain certain (perhaps, empirically unverified) causal
assumptions which go beyond what is fair to assume in the framework of the
mathematical formalism. The first experimental indication on a difference
between propagation velocity of bound and radiation fields can be regarded
as an unexpected causal asymmetry in the structure of possible solutions to
Maxwell's equations. The latter has no grounds in the formalism of the
existing electromagnetic theory and implies that a possible interpretive
framework can be wider than the standard account allows. In our opinion, the
presence of this fundamental uncertainty in the classical EM theory is due
to the fact that perhaps there has not been realized the considerable
importance in careful drawing and empirical testing distinction between
bound and radiation field components. Put in other terms, it has to imply a
considerable shift of attention towards the conception of the velocity
dependent (bound) components of EM field because they may result of a
crucial importance in illuminating alternative foundations of the classical
electrodynamics.

In this respect, advocates of alternative views may appeal to the fact that
at the time of Hertz experiments on propagation of EM interactions there was
already a consistent rival approach to electromagnetic phenomena that
emphasized the distinction between spreading velocity of bound and radiation
EM field components. In the history of physics it is regarded as Helmholtz's
electrodynamics$^{\cite{helm}}$. Moreover, Maxwell's equations appeared in
Helmholtz's theoretical scheme in a limit when the spreading velocity of
bound (electrostatic in Helmholtz's classification) electric components
tended to infinity. In this case, the propagation velocity of radiation
waves (electrodynamic components in Helmholtz's classification) resulted
equal to the velocity of light as it was assumed in Maxwell's theory (for
more details, see$^{\cite{smir},\cite{buchw}}$). Recently, the same
conclusion on a possibility of essential difference between spreading
velocity of bound and radiation EM field components was obtained in
completely independent and mathematically consistent approaches to the
classical electromagnetic theory based on Maxwell's equations$^{\cite{chub1},%
\cite{chub2}-\cite{chub3}}$.

We also have to point out the fact of a nearly perfect coincidence between
experimental data and theoretical predictions based on the Eq. (\ref{ochoa})
for the spreading velocity of bound fields highly exceeding the velocity of
light ($v\gg c$). This circumstance can be considered as a strong indication
on the validity of a new generalized Biot-Savart law in form similar to the
Eq. (\ref{ochoa}) which differs form the generally accepted expression (\ref%
{ocho}) only as regards the spreading velocity of bound components, i.e. the
condition $v\geq c$.

Nevertheless, speaking in strict terms, any discrepancy of experimental
observations with theoretical predictions made on the basis of the standard
condition $v=c$ does not automatically imply the correctness of our
hypothesis in form of the Eq. (\ref{ochoa}) but rather experimentally
observed violation of the applicability of the standard retardation
condition to bound fields. By its significance and importance, this
experimental fact may be comparable with the violation of Bell's
inequalities in quantum mechanics. It is very likely that both
manifestations of non-locality might have the same origin. Moreover, we are
inclined to think that non-local characteristics of bound fields shed a new
promising light on a possible close relationship between classical
electrodynamics and quantum mechanics.

Having in mind the challenge of measuring very feeble effects produced by
bound fields, we have introduced into experimental practice multi-section
emitting and receiving antennas that are designed to rise substantially the
ratio of bound-to-radiation field strength. However, our experimental
experience led us to the conclusion that a direct measurement of the
spreading velocity of bound fields still must be qualified as a rather
difficult task even if one implements experimental facilities available in
modern physics laboratories. In this respect, we would like to emphasize
some technical circumstances of our experimental approach. Due to a rapid
fall of bound field amplitudes as $R^{-2}$ and $R^{-3}$, the range of
variation $\Delta R$ of the distance between emitting and receiving antennas
has been essentially restricted by approximately $1m$. If we consider EM
waves travelling with the velocity of light $c$, the difference $\Delta
R\approx 1m$ means the time shift $\Delta t=\Delta R/c\approx 3ns$. In order
to achieve a necessary precision of measurements (within the time range of $%
3ns$), the front of a generated signal must not exceed the value of $1ns$,
(i.e. $4ns$ for a period of one quasi-harmonic signal) which corresponds to
the angular frequency $\omega =1.5\cdot 10^{9}Hz$. Numerical calculations
show that a typical size of antenna (emitting at the frequency $\omega
=1.5\cdot 10^{9}Hz$) is about several $cm$ and, therefore, at small
distances (for instance, $R\approx 0.1m$) comparable with the size of the
EA, the amplitude of the output signal should be of the order of $1V$. The
crucial parameter that is the ratio of bound-to-radiation field strength
also falls down rapidly with the distance as approximately $\left( \frac{c}{%
\omega R}\right) ^{2}$. In our experiments, this ratio is close to unity at $%
R\approx 0.2m$ and results negligible already at distances exceeding
$1m$. Hence, if there is an intention to produce any observable
separation of bound and radiation components at $R=1m$, the value of
the ratio of bound-to-radiation field strength has to be close to
unity at this distance, i.e. one has to assume that $\omega \approx
c/R=3\cdot 10^{8}Hz$. It implies a considerable rise of the front of
generated signals up to nearly $5ns$ which already will exceed the
permissible time shift of about $3ns$, making impossible any visual
distinction between bound and radiation contributions into the
resultant signal.

Thus, at present stage these circumstances made it almost
unattainable task to undertake direct measurements of spreading
velocity of bound magnetic fields. Having in mind experimental
facilities available at the time of Hertz's experiments (1886-1891),
we might intuitively assume that the problem of propagation of
electromagnetic interactions could hardly have been resolved
definitely at the end of the 19th century in\ favor of one of the
rival theories (Maxwell's or Helmholtz's electrodynamics).
Nevertheless, in contrast to our possible reservations, Hertz's
contemporaries accepted his experiments almost unconditionally, i.e.
by the beginning of the 20th century the problem of propagation of
electromagnetic interactions had been considered as definitely
resolved. It gave a rise to the conviction that the standard
retardation condition ($v=c$) has a general validity for all
components of EM field. In this respect, we stress again that
Hertz's experiments were not especially thought-out to test the
applicability of the retardation condition to bound fields, since at
the time of Hertz's experiments the classical electromagnetism had
not yet been completed in its modern theoretical framework.
Therefore, from the point of view of the modern EM theory, this fact
implies a methodological incompleteness of Hertz's experimental
approach. In our opinion, these historical circumstances combined
with the extreme feebleness of the effects produced by bound fields
deprived the majority of physicists (from the beginning of the 20th
century up till now) of the interest towards experimental study of
bound fields properties.

\section{Acknowledgments}

The authors thank Dr. V. Onoochin (Moscow, Russia) for valuable discussions
and J.A. Santamaria Rueda (Moscow, Russia) for financial support which
resulted crucial in the realization of this project.

\bigskip

\begin{center}
\textbf{APPENDIX. Derivation of the value of magnetic field due to an
idealized magnetic dipole under the general assumption }$v\geq c$
\end{center}

First, it is worth reminding that Eq. (\ref{ocho}) was obtained in$^{\cite%
{jefim}}$ at the low velocity relativistic limit of the exact solution (\ref%
{seis}) of the D'Alembert equation (\ref{siete}) when the centripetal
acceleration of the moving charges is ignored. The latter requirement
implies that in places where the antenna loop is curved, the current flow $%
I(t)$ and the rate of time variation of current $\dot{I}(t)$ are parallel to
the tangential direction along the loop. To illustrate the derivation of Eq.
(\ref{nueve}) from the hypothetical general case driven by Eq. (\ref{ochoa}%
), we shall use the same assumptions as it is done in$^{\cite{jefim}}$ and
follow the analogous procedure implemented in$^{\cite{rosser}}$.

Thus, if the radius $r$ of the loop of the antenna is negligible
in comparison with the distance to the point of observation $R$,
i.e. $r\ll R$, then the shape of the antenna has no importance and
it can be chosen in a way to simplify theoretical calculations.
So, let us consider the small current carrying coil \textit{ABCD}
shown \ in Fig.$9$. The arrow indicates
the direction of the flowing current. The sections \textit{AD} and \textit{BC%
} are arcs circles, with centers at the point of observation $P$. Both
circles have the radius value $R$ and $R+\delta R$ respectively, where $%
\delta R\ll R$. Thus, if the length of the section \textit{AD} is equal to $%
r $ (where $r\ll R$), then the length of \textit{BC} is $r(1+\delta R/R)$.

Let us now determine the magnetic field $\mathbf{B}$ at the point of
observation $P$ and at the time of observation $t$. According to Eq. (\ref%
{ochoa}) the contributions of the currents in the sections \textit{AB} and
\textit{CD} are both zero since $\mathbf{k}\times \mathbf{n}=0$. Thus, we
can calculate the value of the magnetic field generated only by the sections
\textit{BC} and \textit{AD}:

\begin{equation}
\mathbf{B}=\frac{1}{4\pi \varepsilon _{0}c^{2}}\oint\limits_{BC+AD}\left\{
\frac{\left[ \NEG{I}\right] _{v}}{R^{2}}+\frac{\left[ \dot{I}\right] }{cR}%
\right\} \mathbf{k}\times \mathbf{n}dl  \label{A1}
\end{equation}
Integrating Eq. (\ref{A1}) we obtain the value of the magnetic field
generated by the section \textit{BC}:

\begin{equation}
\mathbf{B}_{BC}=\frac{1}{4\pi \varepsilon _{0}c^{2}}\left\{ \left[ \NEG{I}%
\right] _{v}\frac{r(1+\delta R/R)}{R^{2}(1+\delta R/R)^{2}}+\left[ \dot{I}%
\right] \frac{r(1+\delta R/R)}{cR(1+\delta R/R)}\right\} \mathbf{k}
\label{A2}
\end{equation}
and by the section \textit{AD} respectively:

\begin{equation}
\mathbf{B}_{AD}=-\frac{1}{4\pi \varepsilon _{0}c^{2}}\left\{ \left[ \NEG{I}%
\right] _{v}\frac{r}{R^{2}}+\left[ \dot{I}\right] \frac{r}{cR}\right\}
\mathbf{k}  \label{A3}
\end{equation}

Since $\delta R/R\ll 1$, the use of the binomial theorem is justified and
Eq. (\ref{A2}) takes a more compact form:

\begin{equation}
\mathbf{B}_{BC}=\frac{1}{4\pi \varepsilon _{0}c^{2}}\left\{ \left[ \NEG{I}%
\right] _{v}\frac{r}{R^{2}}(1-\frac{\delta R}{R})+\left[ \dot{I}\right]
\frac{r}{cR}\right\} \mathbf{k}  \label{A2a}
\end{equation}

The circumstance that the section \textit{AD }\ is placed closer to the
point of observation $P$ means that any signal travelling with the finite
velocity and produced in this section at some initial instant $t_{0}$ will
arrive at the point $P$ earlier than its \textit{BC} counterpart, generated
at the same time $t_{0}$.

Under the assumption $v\geq c$, bound components of the magnetic fields $%
\mathbf{B}_{BC}$ and $\mathbf{B}_{AD}$ arrive at the point of observation $P$
with different values of retardation time $t^{\prime }=t-(R+\delta R)/v$ and
$t^{\prime }=t-R/v$ respectively. Therefore, both values of current $\left[
\NEG{I}\right] _{v}$ in (\ref{A2}) and (\ref{A3}) have been generated at
different initial instants of time $t_{0}^{BC}=t-(R+\delta R)/v$ and $%
t_{0}^{AD}=t-R/v$, where $t$ is the time of observation at $P$. The shift
between $t_{0}^{BC}$ and $t_{0}^{AD}$ is $\Delta
t_{v}=t_{0}^{AD}-t_{0}^{BC}=\delta R/v$. Both values $I(t_{0}^{BC})$ and $%
I(t_{0}^{AD})$ can be related if one knows the rate of change of current $%
\dot{I}$ in one of the sections:

\begin{equation}
I(t_{0}^{BC})=I(t_{0}^{AD}-\delta R/v)=I(t_{0}^{AD})-\dot{I}%
(t_{0}^{AD})\delta R/v  \label{A4}
\end{equation}

The same reasoning is also applicable to the radiation components which will
arrive at $P$ with different values of retardation time $t^{\prime
}=t-(R+\delta R)/c$ and $t^{\prime }=t-R/c$ respectively. The shift between
corresponding values of retardation time is $\Delta
t_{v=c}=t_{0}^{AD}-t_{0}^{BC}=\delta R/c$. By analogy with (\ref{A4}), it
determines the following relationship:

\begin{equation}
\dot{I}(t_{0}^{BC})=\dot{I}(t_{0}^{AD}-\delta R/v)=\dot{I}(t_{0}^{AD})-\ddot{%
I}(t_{0}^{AD})\delta R/v  \label{A5}
\end{equation}

Thus, Eq. (\ref{A2a}) can be rewritten as:

\begin{equation}
\mathbf{B}_{BC}=\frac{1}{4\pi \varepsilon _{0}c^{2}}\left\{ (\NEG{I}_{0}-%
\dot{I}_{0}\frac{\delta R}{v})\frac{r}{R^{2}}(1-\frac{\delta R}{R})+(\dot{%
\NEG{I}}_{0}-\ddot{I}_{0}\frac{\delta R}{c})\frac{r}{cR}\right\} \mathbf{k}
\label{A6}
\end{equation}
where $I_{0}=I(t_{0}^{AD})=\left[ \NEG{I}\right] _{v}$ and $\dot{I}_{0}=\dot{%
I}(t_{0}^{AD})=\left[ \dot{\NEG{I}}\right] _{v}$\ are evaluated at the
retarded time $t^{\prime }=t-R/v$ whereas $\ddot{I}_{0}=\ddot{I}(t_{0}^{AD})=%
\left[ \ddot{\NEG{I}}\right] $ at $t^{\prime }=t-R/c$.

Adding equations (\ref{A6}) and (\ref{A3}) we find the resultant magnetic
field at the point $P$ and the time $t$:

\begin{equation}
\mathbf{B}=\frac{\Delta S}{4\pi \varepsilon _{0}c^{2}}\left\{ \frac{\left[
\NEG{I}\right] _{v}}{R^{3}}+\frac{c}{v}\frac{\left[ \dot{\NEG{I}}\right] _{v}%
}{cR^{2}}+\frac{\left[ \ddot{\NEG{I}}\right] }{c^{2}R}\right\} \mathbf{k}
\label{A7}
\end{equation}
where we have taken into account that the area of the loop \textit{ABCD} is $%
\Delta S=r\delta R$.

Under the standard retardation constraint $v=c$ valid in respect to all
components of the magnetic field, we arrive at the conventional form$^{\cite%
{rosser}-\cite{stratton}}$:

\begin{equation}
\mathbf{B}(R,t)=-\frac{\Delta S}{4\pi \varepsilon _{0}c^{2}}\left\{ \frac{%
\left[ I\right] }{R^{3}}+\frac{\left[ \dot{I}\right] }{cR^{2}}+\frac{\left[
\ddot{I}\right] }{c^{2}R}\right\} \mathbf{k}=-\frac{\mu _{0}}{4\pi }\left\{
\frac{\left[ \mathbf{m}\right] }{R^{3}}+\frac{\left[ \mathbf{\dot{m}}\right]
}{cR^{2}}+\frac{\left[ \mathbf{\ddot{m}}\right] }{c^{2}R}\right\}
\label{A7a}
\end{equation}
where $\mathbf{m}=I\Delta S\mathbf{k}$ and $\varepsilon _{0}\mu _{0}=1/c^{2}$%
.


\newpage

\begin{figure}[htbp]
\begin{center}
\centerline{\includegraphics[width=5.83in,height=5.83in]{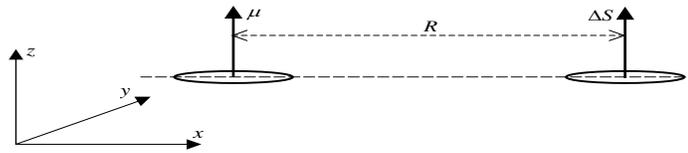}}
\caption{Geometrical configuration of the EA and RA in the plane
$xy$}
\end{center}
\end{figure}

\begin{figure}[htbp]
\begin{center}
\centerline{\includegraphics[width=5.83in,height=5.83in]{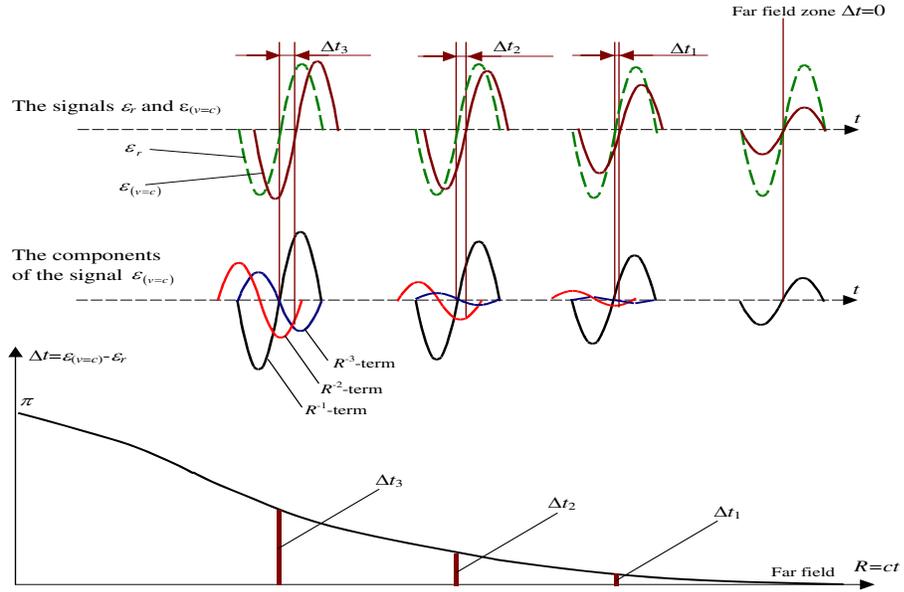}}
\caption{the case $v=c$; evolution (with $R$ and $t$) of the
temporal shift between $\varepsilon _{v=c}(t)$ and $\varepsilon
_{r}(t)$; specification of the relative contribution of $R^{-3}$,
$R^{-2}$ and $R^{-1}$-terms into the resultant signal $\varepsilon
_{v=c}(t)$; dependence $\Delta t(R)$ corresponding to the time
shift between instants when signals $\varepsilon _{v=c}(t)$ and
$\varepsilon _{r}(t)$ cross zero line}
\end{center}
\end{figure}

\begin{figure}[htbp]
\begin{center}
\centerline{\includegraphics[width=5.83in,height=5.83in]{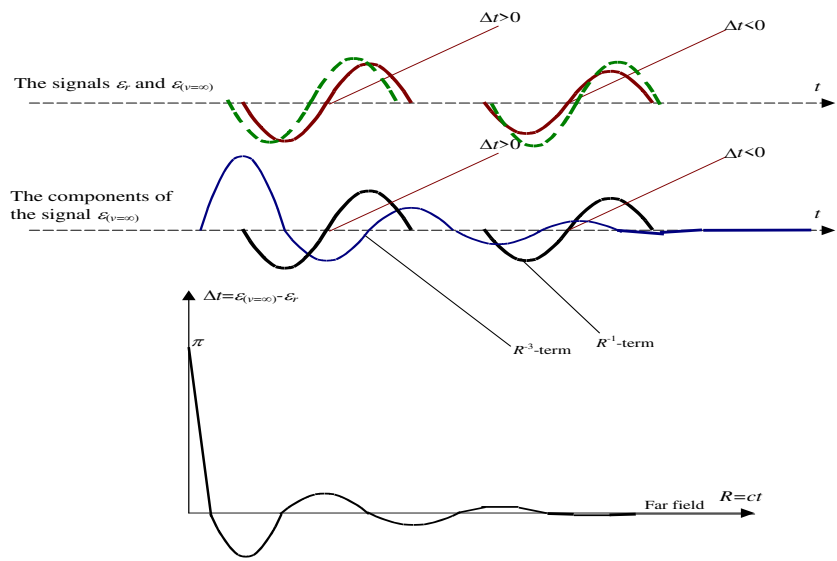}}
\caption{the case $v=\infty$}
\end{center}
\end{figure}

\begin{figure}[htbp]
\begin{center}
\centerline{\includegraphics[width=5.83in,height=5.83in]{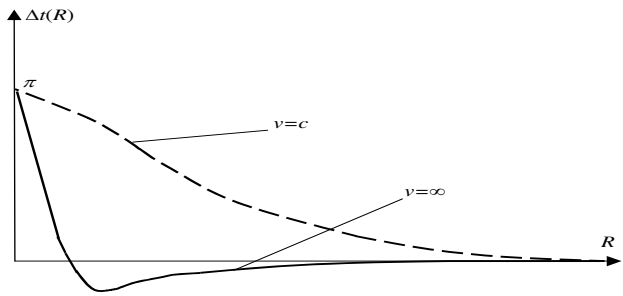}}
\caption{the case $v=\infty $; the dependence $\Delta t(R)$ is
calculated under the assumption that the relative contribution of $R^{-3}$%
-term into the resultant signal $\varepsilon _{v=\infty }(t)$ is
negligible in comparison with $R^{-1}$-term at distances
$R>c/\omega$}
\end{center}
\end{figure}

\begin{figure}[htbp]
\begin{center}
\centerline{\includegraphics[width=5.83in,height=5.83in]{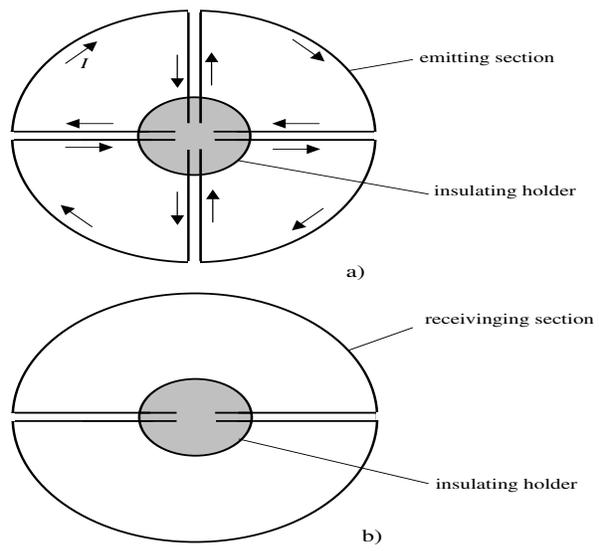}}
\caption{\textit{(a)} cross-section of the EA; the arrows show the
direction of current flows in all sections; \textit{(b)}
cross-section of the RA.}
\end{center}
\end{figure}

\begin{figure}[htbp]
\begin{center}
\centerline{\includegraphics[width=5.83in,height=5.83in]{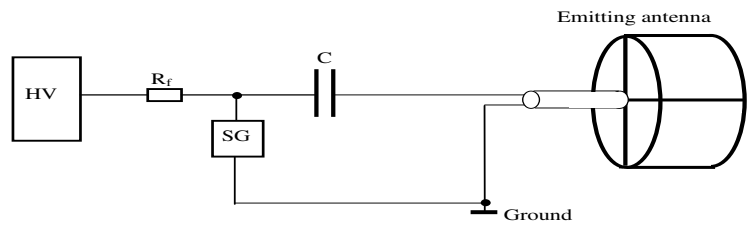}}
\caption{Circuit of the emitting antenna.}
\end{center}
\end{figure}

\begin{figure}[htbp]
\begin{center}
\centerline{\includegraphics[width=5.83in,height=5.83in]{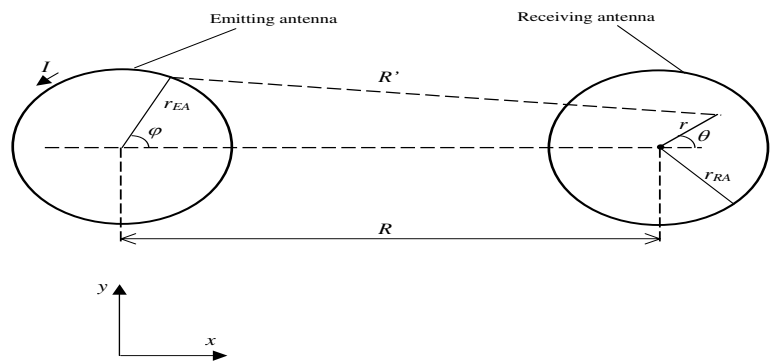}}
\caption{Calculation of the resultant magnetic field $\mathbf{B}$
and e.m.f. $\varepsilon (t)$\ in the loop of the RA; specification
of polar coordinate systems attached to the EA and RA.}
\end{center}
\end{figure}

\begin{figure}[htbp]
\begin{center}
\centerline{\includegraphics[width=5.83in,height=5.83in]{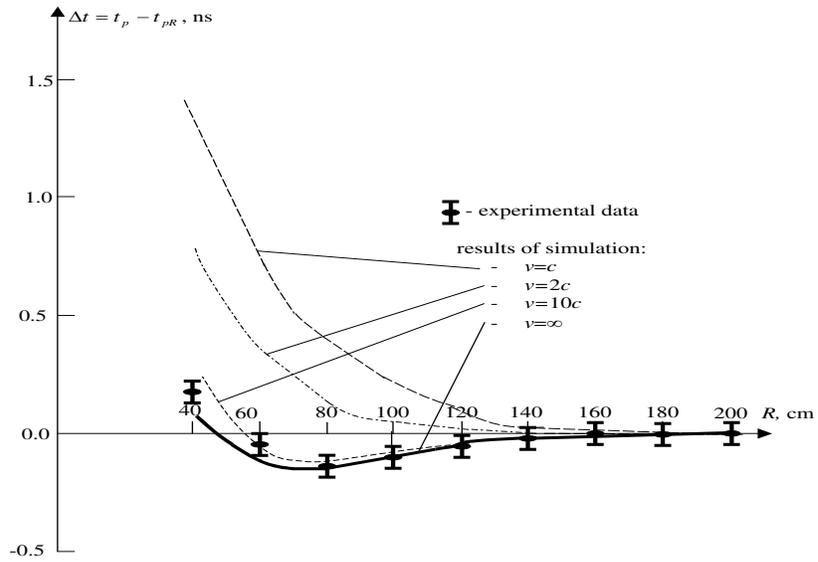}}
\caption{Dot lines illustrate numerical predictions of the
dependence $\Delta t(R)$ for $v=c$; $v=2c$ and $v=10c$; continuous
line corresponds to the numerical predictions of $\Delta t(R)$ for
$v=\infty $; empirically found dependence $\Delta t(R)$\ is
represented by black circles.}
\end{center}
\end{figure}

\begin{figure}[htbp]
\begin{center}
\centerline{\includegraphics[width=5.83in,height=5.83in]{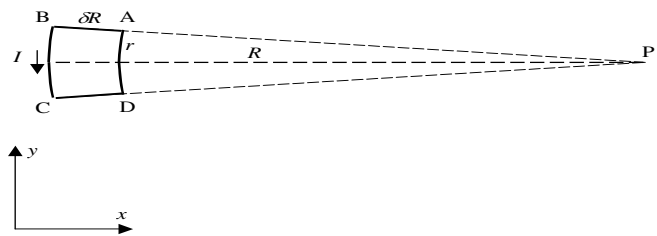}}
\caption{Calculation of the resultant magnetic field $\mathbf{B}$
at the point $P$ due to both bound $\mathbf{B}_{\mathbf{u}}$ and
radiation $\mathbf{B}_{\mathbf{a}}$ components generated by
varying currents $I(t)$ in the loop \textit{ABCD}}
\end{center}
\end{figure}

\end{document}